\documentclass[]{elsarticle}

\usepackage{graphicx}   

\begin{document}
	
\begin{frontmatter}

\title{Generation of a Novel Exactly Solvable Potential}

\author[address1]{Jonathan Bougie}
\ead{jbougie@luc.edu} 

\author[address1]{Asim Gangopadhyaya\corref{mycorrespondingauthor}}
\cortext[mycorrespondingauthor]{Corresponding author}
\ead{agangop@luc.edu} 

\author[address1]{Jeffry V. Mallow}
\ead{jmallow@luc.edu}  

\author[address2]{Constantin Rasinariu}
\ead{crasinariu@colum.edu }   

\address[address1]{Loyola University Chicago, Department of Physics Chicago,
	IL 60660}
\address[address2]{Columbia College Chicago, Department of Science and 
Mathematics, Chicago, IL 60605}

\begin{abstract}
We report a new shape invariant (SI) isospectral extension of the Morse potential. Previous investigations have shown that the list of ``conventional" SI superpotentials that do not depend explicitly on Planck’s constant $\hbar$ is complete. Additionally, a set of ``extended" superpotentials has been identified, each containing a conventional superpotential as a kernel and additional $\hbar$-dependent terms. We use the partial differential equations satisfied by all SI superpotentials to find a SI extension of Morse with novel properties. It has the same eigenenergies as Morse but different asymptotic limits, and does not conform to the standard generating structure for isospectral deformations. 
\end{abstract}


\end{frontmatter}


\section{Introduction}
Supersymmetric quantum mechanics \cite{Witten,Solomonson,CooperFreedman}
generalizes the familiar factorization method for the harmonic oscillator. 
A general hamiltonian $H_-$ is factorized in terms of two operators ${\cal
A}^+ \equiv -\hbar \frac{d}{dx}+ W(x,a)$ and ${\cal A}^- \equiv \hbar
\frac{d}{dx}+ W(x,a)$, where the function $ W(x,a)$ is known as the
superpotential. We have set the mass $m = 1/2$. The product ${\cal A}^+\, {\cal A}^-$ is then given by
\begin{eqnarray}
{\cal A}^+\, {\cal A}^-
&=& \left(  -\hbar \frac{d}{dx}+ W(x,a) \right)  
~\left( \hbar \frac{d}{dx}+ W(x,a)\right) \nonumber \\
&=&  -\hbar^2 \frac{d^2}{dx^2}+W^2(x,a)  -  \hbar \frac{d\, W(x,a)}{dx}~ \label{A+A-},
\end{eqnarray}
 This product ${\cal A}^+\, {\cal A}^-$ produces
a hamiltonian $H_-=-\hbar^2\frac{d^2}{dx^2}+V_-(x,a)$, where $V_-(x,a) =
W^2(x,a) - \hbar \frac{d\, W(x,a)}{dx}$. A product of operators ${\cal A}^-$
and ${\cal A}^+$ carried out in the reverse order produces another
hamiltonian $H_+= -\hbar^2 \frac{d^2}{dx^2}+V_+(x,a)$ with $V_+(x,a) =
W^2(x,a) + \hbar \frac{d\, W(x,a)}{dx}$. The two hamiltonians are
intertwined; \emph{i.e.}, ${\cal A}^+H_+ = H_-{\cal A}^+$ and ${\cal A}^-H_-
= H_+{\cal A}^-$.  This leads to the following isospectrality relationships
among their eigenvalues and eigenfunctions for all integer $n \geq 0 $:
\begin{eqnarray}
E^{-}_{n+1} =E^{+}_{n},
\end{eqnarray}
\begin{eqnarray}
 \frac{~~~{\cal A}^- }{\sqrt{E^{+}_{n} }} ~\psi^{(-)}_{n+1} 
 = ~\psi^{(+)}_{n}  ~{\rm ,and} ~
\frac{~~~{\cal A}^+}{\sqrt{E^{+}_{n} }}~\psi^{(+)}_{n} 
= ~  \psi^{(-)}_{n+1}. \label{isospectrality}
\end{eqnarray}
Eq. (\ref{isospectrality}) implies that if somehow we knew the eigenvalues
and eigenstates of the hamiltonian $H_-$, we would automatically know the
same for the hamiltonian $H_+$, and vice-versa. 

We can go no further unless we know the spectrum for one of the potentials
{\it a priori}. However, if the superpotential $W(x,a_i)$ obeys the
following ``shape invariance condition"
\cite{Infeld,Miller,gendenshtein1,gendenshtein2},
\begin{equation}
W^2(x,a_i)  +  \hbar \frac{d\, W(x,a_i)}{dx}+g(a_i) = 
W^2(x,a_{i+1})  -  \hbar \frac{d\, W(x,a_{i+1})}{dx}+g(a_{i+1}),~ 
\label{SIC1}
\end{equation}
the eigenvalues and eigenfunctions for both hamiltonians can be determined
separately. In this letter we consider the case of translational or additive
shape invariance: $a_{i+1}=a_{i}+\hbar$. From Eq. (\ref{SIC1}) it follows
that for a shape invariant system the partner hamiltonians $H_+(x,a_{i})$
and $H_-(x,a_{i+1})$ differ only by values of parameter $a_i$ and additive
constants $g(a_i)$. The eigenvalues and eigenfunctions of a shape invariant
system are then given
by \cite{Cooper-Khare-Sukhatme,Gangopadhyaya-Mallow-Rasinariu}
\begin{eqnarray}
E_n^{(-)}(a_0)=g(a_n)-g(a_0)~~{\rm
for}~n\geq0~,\nonumber
\label{energy_sic} 
\end{eqnarray}
and
\[
\psi^{(-)}_{n}(x,a_0)= 
\frac{{\cal A}^+{(a_0)} 
	~ {\cal A}^+{(a_1)}  \cdots  {\cal A}^+{(a_{n-1})} ~ \psi^{(-)}_0(x,a_n)}
	  {\sqrt{E_{n}^{(-)}(a_0)\,E_{n-1}^{(-)}(a_1)\cdots E_{1}^{(-)}(a_{n-1})}}
~,
\]
where $\psi^{(-)}_0(x,a_n) = N e^{-\frac1\hbar\, \int^x W(y,a_n)\,dy}$ is the solution of ${\cal A}^-\psi^{(-)}_0 =0 $ and $N$ is the normalization constant.
Thus, all additive shape invariant systems are exactly solvable. Therefore,
finding all such potentials gives important insight into the nature of
quantum mechanical systems. Previously, researchers found a list of additive
shape invariant potentials, generally by trial and error
\cite{Infeld,Dutt,CGK}. In Ref. \cite{Bougie2010,symmetry} the authors
showed that all ``conventional'' ($\hbar$-independent) shape invariant
superpotentials could be determined as solutions to a set of two partial
differential equations. Additionally, they showed that the list of
conventional shape invariant superpotentials given in Ref.\cite{Dutt} was
indeed complete.

A new class of ``extended'' superpotentials was discovered
\cite{Quesne1,Quesne2} and expanded upon \cite{Odake1, Odake2, Tanaka,
Odake3, Odake4, Quesne2012a, Quesne2012b, Ranjani1,Ranjani2}. These
superpotentials contain explicit $\hbar$-dependence in $W$.  When expanded
in powers of $\hbar$, each power of $\hbar$ yields a new partial
differential equation that must be satisfied by these superpotentials
\cite{Bougie2010,symmetry}.  

Although this formalism has been checked for consistency with already known 
superpotentials \cite{Bougie2010,symmetry}, it has not been used to generate
a new superpotential.  In this letter we employ it to generate an extension
of the Morse superpotential. This extended superpotential exhibits 
properties that are qualitatively different from those previously
discovered.  It has different asymptotic values than Morse and it contains
free parameters. 

\section{A Novel Potential}
We first consider conventional superpotentials that do not have an intrinsic
dependence on $\hbar$.  Note that in Eq. (\ref{SIC1}), for additive shape
invariance the constant $\hbar$ enters $W$ through the linear combination
$a+ \hbar$; therefore, 
\mbox{$ \frac{\partial}{\partial \hbar} W(x,a+\hbar)
= \frac{\partial}{\partial a} W(x,a+ \hbar)$}.
Since Eq. (\ref{SIC1}) must hold for an arbitrary value of $\hbar$, we can
expand it in powers of $\hbar$, and stipulate that the coefficient of each
power vanishes. This leads to two independent equations
\cite{Bougie2010,symmetry}:
\begin{eqnarray}\vspace{6pt} W \, \frac{\partial W}{\partial a} 
- \frac{\partial W}{\partial x} + \frac12 \, \frac{d g(a)}{d a} = 0~
\label{PDE1}
\end{eqnarray} and
\begin{eqnarray}
\frac{\partial^{3}}{\partial a^{2}\partial x} ~W(x,a)= 0\label{PDE3}~.
\end{eqnarray}
The general solution to Eq. (\ref{PDE3}) is
\begin{eqnarray}
W(x,a)=a\cdot {X}_1(x)+{X}_2(x)+u(a)~\label{GeneralSolution},
\end{eqnarray}
where $X_1, X_2$, and $u$ are arbitrary functions. When combined with Eq. (\ref{PDE1}), this general solution reproduces the
complete family of conventional superpotentials, as shown in Table
\ref{Table1} \cite{Bougie2010,symmetry}.
\begin{table} [htb]
\begin{center}
\begin{tabular}{l l}
\hline
Name  &  Superpotential   \\
\hline
Harmonic Oscillator &  $\frac12 \omega x$  \\

Coulomb   &  $\frac{e^2}{2(\ell+1)} - \frac{\ell+1}{r}$\\

3-D oscillator  & $\frac12 \omega r - \frac{\ell+1}{r}$ \\

Morse &$A-Be^{-x}$  \\

Rosen-Morse I &$-A\cot x - \frac{B}{A}$  \\

Rosen-Morse II &$A\tanh x + \frac{B}{A}$  \\

Eckart &$-A\coth x + \frac{B}{A}$  \\

{Scarf I} &  $ A\tan x - B {\rm sec}\,x$ \\

{Scarf II}  &  $A\tanh x + B {\rm sech}\,x$\\

Gen. P\"oschl-Teller &  $A\coth x- B {\rm cosech}\,x $  \\
\hline
\end{tabular}
\caption{The complete family of $\hbar$-independent additive shape-invariant 
superpotentials.}
\label{Table1}
\end{center}
\end{table}

The additional set of extended superpotentials reported in
\cite{Quesne1,Quesne2} obey the shape invariance condition given in
Eq.~(\ref{SIC1}) only when their $W$'s depend explicitly on $\hbar$. Each of
these extended superpotentials is isospectral with one of the conventional
superpotentials listed in Table \ref{Table1}.  

To determine the differential equations obeyed by extended superpotentials,
we expand the superpotentials in powers of $\hbar$.  I.e., we define
\begin{eqnarray}
\vspace{6pt}
    W(x, a, \hbar) = \sum_{j=0}^\infty \hbar^j W_j(x,a).~ \label{W-hbar}
\end{eqnarray}
Since $a_0=a$ and $a_1=a+\hbar$,  this power series yields 
\begin{eqnarray}
W\left(x,a_1,\hbar\right)=\sum_{j=0}^\infty \sum_{k=0}^{j}\frac{\hbar^j}{k!}
\frac{\partial^k W_{j-k}}{\partial a^k},
\nonumber
\end{eqnarray} 
\begin{eqnarray}
W^2\left(x,a_1,\hbar\right)=
 \sum_{j=0}^{\infty}\sum_{s=0}^{j}\sum_{k=0}^{s}
 \frac{\hbar^j}{\left(j-s\right)!}\frac{\partial^{j-s}
\left(W_{k}W_{s-k}\right)}{\partial a^{j-s}},
\nonumber
\end{eqnarray}
and
\begin{eqnarray}
\left.\frac{\partial W}{\partial x}\right|_{a=a_1}=
\sum_{j=0}^{\infty}\sum_{k=0}^{j}\frac{\hbar^j}{k!}
\frac{\partial^{k+1}W_{j-k}}{\partial a^k \partial x}
\nonumber~~.
\end{eqnarray}
We then substitute these into Eq. (\ref{SIC1}) and demand that the equation
hold for any value of $\hbar$.  This is possible only if the coefficients of
the series for each power of $\hbar$ are identically zero.  We obtain for
$j=1$
\begin{eqnarray}
2\frac{\partial W_{0}}{\partial x}
-\frac{\partial }{\partial a} \left(W_{0}^2+g  \right) = 0,
\label{lowestorder}
\end{eqnarray}
for $j = 2$
\begin{eqnarray}
\frac{\partial W_{1}}{\partial x}
-
\frac{\partial}{\partial a} \left( W_0\, W_{1}\right)  = 0,
\label{secondorders}
\end{eqnarray}
and for $j \geq 3$
\begin{eqnarray}
2\,\frac{\partial W_{j-1}}{\partial x}
-\sum_{s=1}^{j-1} \sum_{k=0}^s \frac{1}{(j-s)!}
\frac{\partial^{j-s}}{\partial a^{j-s}} W_k\, W_{s-k}+
\nonumber \\
\sum_{k=2}^{j-1} \frac{1}{(k-1)!}
\frac{\partial^{k}\, W_{j-k}}{\partial a^{k-1} \,\partial x} 
+ \left( \frac{j-2}{j!}\right)\,
\frac{\partial^{j} W_0}{\partial a^{j-1}\partial x}   
=0~.
\label{higherorders}
\end{eqnarray}
These equations appear to differ from similar equations in Ref.
\cite{Bougie2010,symmetry}, but they are in fact equivalent. 

We first observe that Eq. (\ref{lowestorder}) is identical to Eq.
(\ref{PDE1}), i.e., all conventional shape invariant potentials
automatically satisfy Eq. (\ref{lowestorder}).  Hence, conventional
superpotentials can be used as kernels to build extended superpotentials. If
$W_0$ is a conventional superpotential, then Eq. (\ref{PDE3}) is
automatically satisfied for $W_0$, meaning that the last term of Eq.
(\ref{higherorders}) disappears.

In the remainder of this paper, we use the Morse superpotential as a kernel
to construct an extended superpotential. In Ref. \cite{Quesne2012b}, the
author generated a quasi-exactly solvable extension of Morse and showed that
it was not shape invariant. 

The strength of the method used in this letter is that we can employ Eq.
(\ref{higherorders}) term-by-term in order to generate a manifestly
shape-invariant solution. Hence, we begin with $W_0=(\alpha-a) - e^{-x}$ and
choosing $W_1=0$, the equation for $W_2$ reads
\begin{eqnarray}
\frac{\partial W_{2}}{\partial x}
- \frac{\partial W_{0} W_2}{\partial a}  = 0.\nonumber 
\label{Eq_order3}
\end{eqnarray}
This equation is solved by
\begin{eqnarray}W_{2}(x,a) =e^{- x} \left( 2 P + Q e^{-2 x} 
- 2 \left(\alpha - a\right) Q e^{- x}\right)~,
\nonumber\end{eqnarray}
where $P$ and $Q$ are constant parameters.
Assuming $W_3$ to be zero, the equation for $W_4$ is 
\begin{eqnarray}
2\frac{\partial W_4}{\partial x}- 
2\frac{\partial \left( W_0 W_4+\frac12 W_2^2\right) }{\partial a}  - 
\frac{\partial }{\partial a}
\left( \frac{\partial W_{2}}{\partial a} 
\frac{\partial W_{0}}{\partial a}  \right) - \nonumber \\
 \frac13 \left(
W_2 \frac{\partial^3 W_{0}}{\partial a^3} 
+
W_0 \frac{\partial^3 W_{2}}{\partial a^3} \right) 
+ \frac{1}{2}\frac{\partial^3 W_{2}}{\partial x \partial a^2}
= 0
\label{Eq_order5}~.
\nonumber\end{eqnarray}
The above equation is solved by  
\begin{eqnarray}
W_{4}(x,a) = - Q e^{-3 x} \left( 2 P + Q e^{-2 x} 
- 2 \left(\alpha - a\right)  Q e^{- x}\right)~.
\nonumber\end{eqnarray}
Generalizing this process yields  
$W_{2k-1}=0$ and 
\begin{equation}
W_{2k}= 
\left(- Q\right)^{k-1} e^{-(2k-1) x} \left( 2 P + Q e^{-2 x} 
- 2 \left(\alpha - a \right)  Q e^{- x}\right)\nonumber
\end{equation}
for all positive integers $k.$ 
Computing the infinite sum $\sum_{j=0}^\infty \hbar^j 
W_j(x,a)$, 
we obtain 
\begin{equation}
W(x,a,\hbar) = (\alpha-a)-e^{- x}  +
\frac{\hbar^2 \left( 2 Pe^{x} - 2 \left( \alpha-a\right)  Q + Q e^{- x} 
\right)}{e^{2 x}+Q\,\hbar^2}~.
\label{newpotential}\end{equation}

\begin{figure}[!htbp]
\centering
\includegraphics[width=0.75\linewidth]{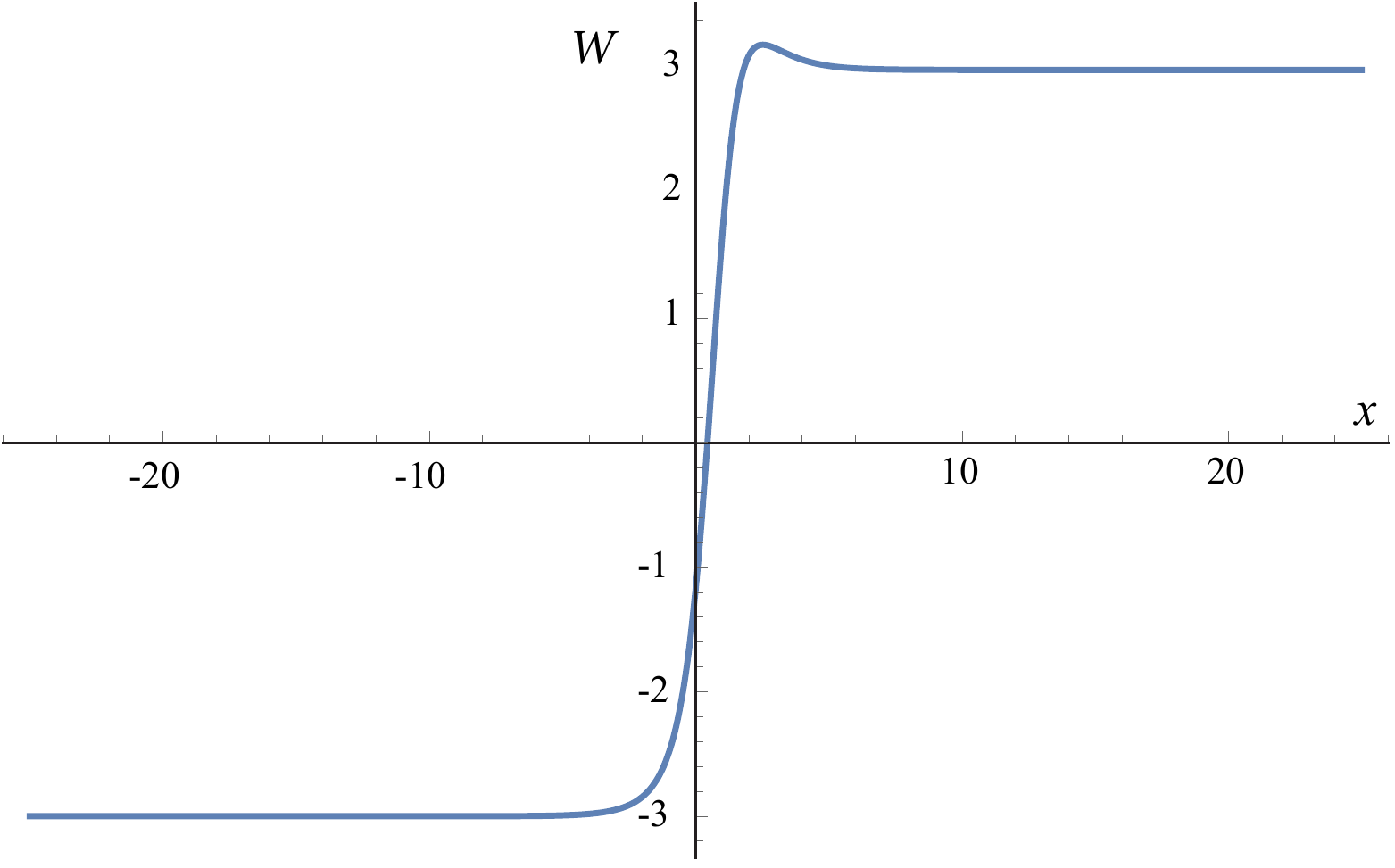}
\caption{Superpotential $W$ for $P=3, Q =5, \alpha=5, a = 2$, and $ 
\hbar = 1.$ Its asymptotic values imply an unbroken supersymmetry. }
\label{fig:MorseExt1}
\end{figure}

\begin{figure}[htb]
	\centering
	\includegraphics[width=0.75\linewidth]{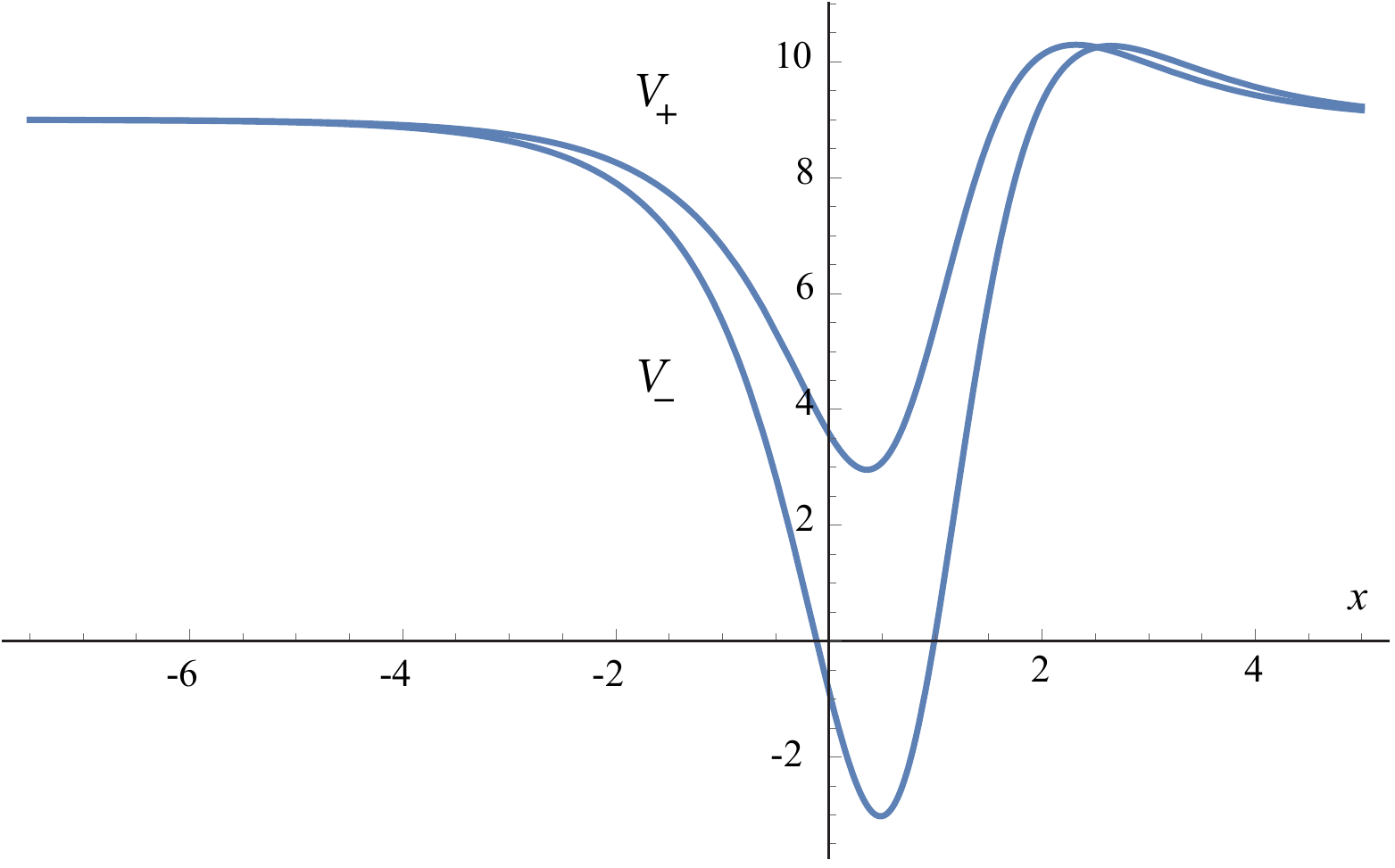}
	\caption{Potentials $V_-$ and $V_+$ for the same values of 
	the parameters as Fig. 1.  The deeper potential is $V_-$, 
	and it holds the zero-energy groundstate.}
	\label{fig:MorseExt4}
\end{figure}

\noindent 
The shape invariance of a superpotential of this form can be directly
checked. Substituting the above expression into Eq. (\ref{SIC1}) yields
\begin{equation}
W^2(x,a) - W^2(x,a+\hbar)  +
 \hbar \frac{d\,}{dx} \left( W(x,a)+W(x,a+\hbar)\right) 
=
 \hbar (2 (\alpha-a) - \hbar),
\end{equation}
which can be brought into the form of Eq. (\ref{SIC1}) by choosing $g(a) =
-(\alpha-a)^2$.  This leads to the energy eigenvalues $E^{(-)}_n =
g(a+n\hbar)-g(a) = (\alpha-a)^2 - (\alpha-a-n\hbar)^2$. These values are
exactly the same as those of the Morse potential.  Figs.
(\ref{fig:MorseExt1}) and (\ref{fig:MorseExt4}) show $W$ and the partner
potentials $V_-$ and $V_+$, where we have chosen the specific values $A=3, B
=5, \alpha=5, a = 2$, and $\hbar = 1 $, as an example.  Asymptotic limits
for this extended superpotential are finite and their absolute values are
equal. Consequently, the resulting potential has identical limits at $\pm
\infty$. 

The ground state eigenfunction can be obtained from the first order 
differential equation $\mathcal{A}^-(x,a)\, \psi_0(x,a) = 0$, yielding
\[
\psi_0(x,a)=  N \left(\hbar^2 Q+e^{2 x}\right)^{a-\alpha}
 \exp \left[x (\alpha-a)+\frac{\left(1-2 \hbar^2 P\right) 
 \tan ^{-1}\left(\frac{e^x}{\hbar \sqrt{Q}}\right)}{\hbar \sqrt{Q}}\right] ,
\]
where $N$ is the normalization constant. The excited states can be obtained
recursively by applying the $\mathcal{A}^+$ operator to $\psi_0$. The first
excited state is $\psi_1(x,a)= \mathcal{A}^+(x,a) 
\psi_0(x,a+\hbar)/\sqrt{E_1}$, and so on. For example, with our particular
choice of parameters, the potential $V_-(x)$ holds three bound states,
corresponding to the eigenenergies $0,5$ and $8$. The fourth energy level is
at $9$, which does not hold a bound state. We obtain for the ground state
eigenfunction 
\[
\psi_0(x)= \frac{40}{\left(e^{2 x}+5\right)^3} 
\sqrt{\frac{105}{1+e^{-\sqrt{5} \pi }}}
\exp\left[
3 x-\sqrt{5} \tan ^{-1}\left(\frac{e^x}{\sqrt{5}}\right)
\right] ,
\]
and for the first two excited states  
\[
\psi_1(x)= \frac{20 \left(\,e^{2 x} + 2 e^x -5\right) \sqrt{\frac{70}{1+e^{\sqrt{5} \pi }}}\, 
} {\left(e^{2 x}+5\right)^3} 
\exp\left[{2 x-\sqrt{5} \tan ^{-1} 
\left(\frac{e^x}{\sqrt{5}}\right)+\frac{\sqrt{5} \pi }{2}}\right] ,
\nonumber
\]
\begin{eqnarray}
\psi_2(x)=&&\frac{4 \left(3 e^{4 x}+20 e^{3 
		x}-20 e^{2x}-100 e^x+75\right) \sqrt{\frac{5}{1+e^{\sqrt{5} \pi }}}}{\left(e^{2 
x}+5\right)^3} \nonumber \\
&& \times \exp \left[x-\sqrt{5} 
\tan ^{-1}\left(\frac{e^x}{\sqrt{5}}\right)+\frac{\sqrt{5} \pi 
}{2} \right] . \nonumber
\end{eqnarray}

\begin{figure}[htb]
	\centering
	\includegraphics[width=0.75\linewidth]{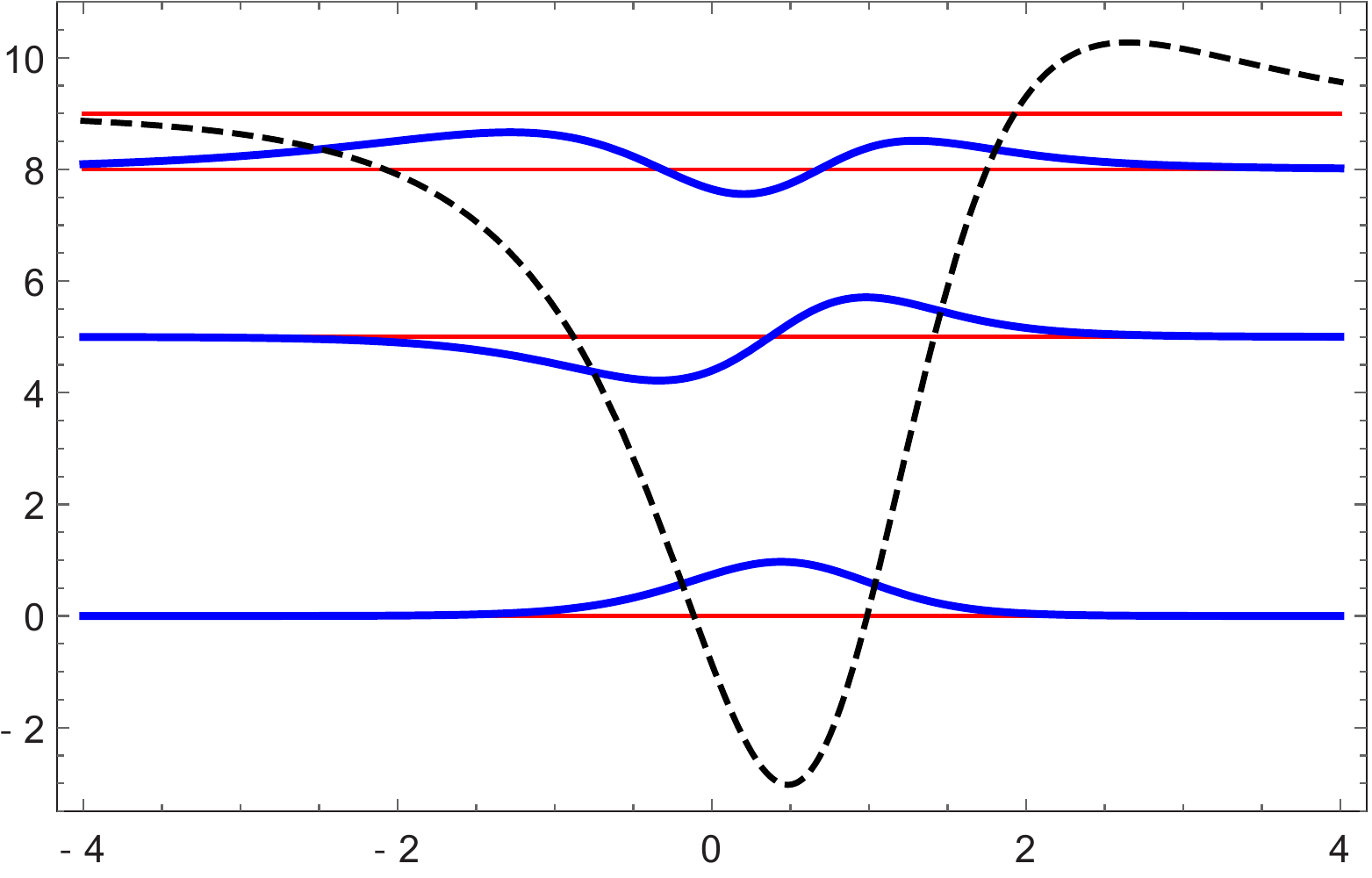}
	\caption{The three bound-state eigenfunctions (shown in blue) for $V_-$ 
	(dashed line) with $P=3, Q =5, 	\alpha= 5, a = 2$, and $ \hbar = 1.$ 
	Each eigenfunction is shown superposed on its corresponding eigenenergy 
	(shown as a horizontal red line). 
	}
	\label{fig:eigen}
\end{figure}

As expected, the fourth energy level does not hold a bound state, as its
corresponding eigenfunction is not square integrable.  Fig.
(\ref{fig:eigen}) superimposes over the potential $V_-(x)$ the three
bound-state eigenfunctions together with their corresponding eigenenergies.

As $\hbar \rightarrow 0$, this transforms into the Morse potential, as
expected. In addition, parameters $P$ and $Q$ also allow us to deform the
potential away from the Morse potential without affecting its eigenvalues.
In this sense, we can consider it to be a shape-invariant isospectral
deformation of Morse.

\section{Conclusions}
In this letter, we have shown that the partial differential equations
(\ref{higherorders}) can be used to generate a shape invariant extension of
the Morse superpotential, whose asymptotic limits differ from those of
Morse. In addition, the free parameters available in this model allow for
its isospectral deformation away from Morse.


\section*{References}

\begin{thebibliography}{10}
\expandafter\ifx\csname url\endcsname\relax
  \def\url#1{\texttt{#1}}\fi
\expandafter\ifx\csname urlprefix\endcsname\relax\def\urlprefix{URL }\fi
\expandafter\ifx\csname href\endcsname\relax
  \def\href#1#2{#2} \def\path#1{#1}\fi

\bibitem{Witten}
E.~Witten, Dynamical breaking of supersymmetry, Nucl. Phys. B185 (1981)
  513--554.

\bibitem{Solomonson}
P.~Solomonson, J.~W. Van~Holten, Fermionic coordinates and supersymmetry in
  quantum mechanics, Nucl. Phys. B196 (1982) 509--531.

\bibitem{CooperFreedman}
F.~Cooper, B.~Freedman, Aspects of supersymmetric quantum mechanics, Ann. Phys.
  146 (1983) 262--288.

\bibitem{Infeld}
L.~Infeld, T.~E. Hull, The factorization method, Rev. Mod. Phys. 23 (1951)
  21--68.

\bibitem{Miller}
W.~Miller~Jr, Lie Theory and Special Functions (Mathematics in Science and
  Engineering), Accademic Press, New York, NY, USA, 1968.

\bibitem{gendenshtein1}
L.~E. Gendenshtein, Derivation of exact spectra of the schrodinger equation by
  means of supersymmetry, JETP Lett. 38 (1983) 356--359.

\bibitem{gendenshtein2}
L.~E. Gendenshtein, I.~V. Krive, Supersymmetry in quantum mechanics, Sov. Phys.
  Usp. 28 (1985) 645--666.

\bibitem{Cooper-Khare-Sukhatme}
F.~Cooper, A.~Khare, U.~Sukhatme, Supersymmetry in Quantum Mechanics, World
  Scientific, Singapore, 2001.

\bibitem{Gangopadhyaya-Mallow-Rasinariu}
A.~Gangopadhyaya, J.~Mallow, C.~Rasinariu, Supersymmetric Quantum Mechanics: An
  Introduction, World Scientific, Singapore, 2010.

\bibitem{Dutt}
R.~Dutt, A.~Khare, U.~Sukhatme, Supersymmetry, shape invariance and exactly
  solvable potentials, Am. J. Phys. 56 (1988) 163--168.

\bibitem{CGK}
F.~Cooper, J.~Ginocchio, A.~Khare, Relationship between supersymmetry and
  solvable potentials, Phys. Rev. D 36 (1987) 2458--2473.

\bibitem{Bougie2010}
J.~Bougie, A.~Gangopadhyaya, J.~V. Mallow, Generation of a complete set of
  additive shape-invariant potentials from an euler equation, Phys. Rev. Lett.
  (2010) 210402:1--210402:4.

\bibitem{symmetry}
J.~Bougie, A.~Gangopadhyaya, J.~V. Mallow, C.~Rasinariu, Supersymmetric quantum
  mechanics and solvable models, Symmetry 4~(3) (2012) 452--473.

\bibitem{Quesne1}
C.~Quesne, Exceptional orthogonal polynomials, exactly solvable potentials and
  supersymmetry, J. Phys. A 41 (2008) 392001:1--392001:6.

\bibitem{Quesne2}
C.~Quesne, Solvable rational potentials and exceptional orthogonal polynomials
  in supersymmetric quantum mechanics, Sigma 5 (2009) 084:1--084:24.

\bibitem{Odake1}
S.~Odake, R.~Sasaki, Infinitely many shape invariant discrete quantum
  mechanical systems and new exceptional orthogonal polynomials related to the
  wilson and askey-wilson polynomials, Phys. Lett. B 682 (2009) 130--136.

\bibitem{Odake2}
S.~Odake, R.~Sasaki, Another set of infinitely many exceptional $(x_\ell)$
  laguerre polynomials, Phys. Lett. B 684 (2010) 173--176.

\bibitem{Tanaka}
T.~Tanaka, N-fold supersymmetry and quasi-solvability associated with
  x-2-laguerre polynomials, J. Math. Phys. 51 (2010) 032101:1--032101:20.

\bibitem{Odake3}
S.~Odake, R.~Sasaki, Exactly solvable quantum mechanics and infinite families
  of multi-indexed orthogonal polynomials, Phys. Lett. B 702 (2011) 164--170.

\bibitem{Odake4}
R.~Odake, S.and~Sasaki, Extensions of solvable potentials with finitely many
  discrete eigenstates, J. Phys. A 46 (2013) 235205.

\bibitem{Quesne2012a}
C.~Quesne, Novel enlarged shape invariance property and exactly solvable
  rational extensions of the rosen-morse ii and eckart potentials, Sigma 8
  (2012) 080.

\bibitem{Quesne2012b}
C.~Quesne, Revisiting (quasi-)exactly solvable rational extensions of the morse
  potential, Int. J. Mod. Phys. A 27 (2012) 1250073.

\bibitem{Ranjani1}
S.~Sree~Ranjani, P.~K. Panigrahi, A.~K. Kapoor, A.~Khare, A.~Gangopadhyaya,
  Exceptional orthogonal polynomials, qhj formalism and swkb quantization
  condition, Jour. of Phys. A 45 (2012) 055210.

\bibitem{Ranjani2}
S.~Sree~Ranjani, R.~Sandhya, A.~K. Kapoor, Shape invariant rational extensions
  and potentials related to exceptional polynomials, arXiv:1503.01394.

\end{thebibliography}
\end{document}